\documentclass[showpacs,amsmath,amssymb,aps,showkeys,floatfix,prd,a4paper]{revtex4}

\usepackage[dvips]{graphicx}
\usepackage{dcolumn}
\usepackage{bm}
\usepackage{epsfig}
\usepackage{amsfonts}
\usepackage{amssymb,amscd}

\def\lsim{\raise0.3ex\hbox{$<$\kern-0.75em\raise-1.1ex\hbox{$\sim$}}}

\def\gsim{\raise0.3ex\hbox{$>$\kern-0.75em\raise-1.1ex\hbox{$\sim$}}}

\def\pom{{I\!\!P}}

\newcommand{\be}{\begin{equation}}

\newcommand{\ee}{\end{equation}}

\def\beq{\begin{equation}}

\def\eeq{\end{equation}}

\def\beqa{\begin{eqnarray}}

\def\eeqa{\end{eqnarray}}

\newcommand{\ba}{\begin{eqnarray}}

\newcommand{\tdm}[1]{\mbox{\boldmath $#1$}}

\newcommand{\rr}{\mbox{\boldmath $r$}}

\newcommand{\rb}{\mbox{\boldmath $b$}}

\def\gappeq{\mathrel{\rlap {\raise.5ex\hbox{$>$}}

{\lower.5ex\hbox{$\sim$}}}}

\def\lappeq{\mathrel{\rlap{\raise.5ex\hbox{$<$}}

{\lower.5ex\hbox{$\sim$}}}}

\def\Toprel#1\over#2{\mathrel{\mathop{#2}\limits^{#1}}}

\def\pom{{I\!\!P}}

\begin{document}

\title{Non-linear QCD dynamics in two-photon interactions at high energies}
\author{ V.P. Gon\c{c}alves $^{1}$, M.S. Kugeratski $^{2}$, E.R. Cazaroto$^{3}$, 
F. Carvalho$^{4}$  and  F.S. Navarra$^3$}
\affiliation{$^{1}$ Instituto de F\'{\i}sica e Matem\'atica,  Universidade
Federal de Pelotas\\
Caixa Postal 354,  96010-900, Pelotas, RS, Brazil.\\
$^2$ Centro de Engenharia da Mobilidade, Universidade Federal de Santa Catarina,\\
Campus Universit\'ario, Bairro Bom Retiro  89219-905, Joinville, SC, Brazil.\\
$^3$Instituto de F\'{\i}sica, Universidade de S\~{a}o Paulo,
C.P. 66318,  05315-970 S\~{a}o Paulo, SP, Brazil\\
$^4$ Departamento de Ci\^encias Exatas e da Terra, Universidade Federal de S\~ao Paulo,\\  
Campus Diadema, Rua Prof. Artur Riedel, 275, Jd. Eldorado, 09972-270, Diadema, SP, Brazil
}

\begin{abstract}

Perturbative QCD predicts that the growth of the gluon density at high energies should saturate, forming a  Color Glass Condensate (CGC), which is described in mean field approximation by the Balitsky-Kovchegov (BK) equation. In this paper we study the $\gamma \gamma$ interactions at high energies and estimate the main observables which will be probed at future linear colliders using the color dipole picture. We discuss in detail the dipole - dipole cross section and propose a new relation between this quantity and the dipole scattering amplitude. The 
total $\gamma \gamma$, $\gamma^{*} \gamma^{*}$ 
cross-sections and  the real photon structure function $F_2^{\gamma}(x,Q^2)$ are calculated using the  recent solution of 
the BK equation with running coupling constant and the predictions are compared with those obtained using phenomenological 
models for the dipole-dipole cross section and scattering amplitude.  We demonstrate that these models are able to describe the 
LEP data at high energies, but predict a very  different behavior for the observables at higher energies. Therefore we conclude 
that the study of   $\gamma \gamma$ interactions can be useful to constrain the QCD dynamics.

\end{abstract}



\maketitle

\vspace{1cm}

\section{Introduction}

The high energy limit of  perturbative QCD is characterized by a center-of-mass energy
 which is much larger than the  hard scales present in the problem. The simplest process
where   this limit can be studied  is the high energy scattering
between two heavy quark-antiquark states, {\it i.e.} the
onium-onium  scattering. For a sufficiently heavy onium state,
high energy scattering is  a perturbative process since the onium
radius gives the essential scale at which the running coupling
$\alpha_s$ is evaluated. In the dipole picture \cite{dipole},
the heavy quark-antiquark pair and the soft gluons in the limit of
large number of colors $N_c$ are viewed as a collection of color
dipoles. In this case, the cross section can be understood as a
product of the number of dipoles in one onium state, the
number of dipoles in the other onium state and  the  basic cross
section for dipole-dipole scattering. At
leading order (LO),  the cross section grows rapidly with the
energy ($\sigma \propto \alpha_s^2  \, e^{(\alpha_{\pom} - 1)Y}$, where
$(\alpha_{\pom} - 1) = \frac{4\alpha_s\,N_c}{\pi}\,\ln 2 \approx 0.5$ and $Y =
\ln\,s/Q^2$) because the LO BFKL equation \cite{bfkl} predicts that the number of dipoles in the light cone
wave function grows rapidly with the energy. Several shortcomings are present in this calculation. Firstly, 
in the  leading order calculation  the energy scale is arbitrary, which implies that the absolute value of  
the total cross section is therefore not predictable. Secondly, $\alpha_s$ is not running at LO BFKL. Finally, 
the power growth with energy violates $s$-channel unitarity at large energies. Consequently, new physical 
effects should modify the LO BFKL equation at very large $s$, making the resulting amplitude unitary.

A theoretical possibility to modify this behavior in a way consistent with the unitarity is the idea of parton saturation,  
where non-linear effects associated to high parton density are taken into account. The basic idea is that when the parton 
density increases (and the scattering amplitude tends to   the unitarity limit), the  linear description present in the BFKL 
equation breaks down and one enters the saturation regime.  
In this regime, the growth of the parton
distribution is expected to saturate, forming a  Color Glass Condensate (CGC), whose evolution with energy is described by an 
infinite hierarchy of coupled equations for the correlators of  Wilson lines (For recent reviews see \cite{cgc}).  
In the mean field approximation, the first equation of this  hierarchy decouples and boils down to a single non-linear integro-differential  equation: the Balitsky-Kovchegov (BK) equation. In the last years 
the next-to-leading order corrections to the  BK equation were calculated  
\cite{kovwei1,javier_kov,balnlo} through the ressumation of $\alpha_s N_f$ contributions to 
all orders, where $N_f$ is the number of flavors. Such calculation allows one to estimate 
the soft gluon emission and running coupling corrections to the evolution kernel.
The authors have found that  the dominant contributions come from the running 
coupling corrections, which allow us to  determine the scale of the running coupling in the 
kernel. The solution of the improved BK equation was studied in detail in Ref. 
\cite{javier_kov}.  In \cite{bkrunning} a global 
analysis of the small $x$ data for the proton structure function using the improved BK 
equation was performed  (See also Ref. \cite{weigert}). In contrast to the  BK  equation 
at leading logarithmic $\alpha_s \ln (1/x)$ approximation, which  fails to describe the HERA 
data, the inclusion of running coupling effects in the evolution renders the BK equation 
compatible with them (See also \cite{vic_joao,alba_marquet,vicmagane}).

A reaction which is analogous to  the process of scattering of two onia discussed above is the 
off-shell photon scattering at high energy in $e^+\,e^-$ colliders, where
the photons are produced from the lepton beams by bremsstrahlung (For a review see, e.g., Ref. 
\cite{nisius}). In these two-photon reactions, the photon virtualities can be made
large enough to ensure the applicability of  perturbative
methods or can be varied in order to test the transition between the soft and hard regimes of the QCD dynamics. 
From the point of  view of the BFKL approach, there are several calculations using the leading
logarithmic  approximation \cite{bartels,gamboone}  and
considering some of the next-to-leading corrections to the total $\gamma^* \gamma^*$
cross section \cite{gamboone,gammaNLO,victor_sauter,papa}. 
On the other hand,  the successful description of all
inclusive and diffractive deep inelastic data from  HERA
by saturation models \cite{GBW,bgbk,kw,kmw,fss04,fss06,kkt,iim,dhj,gkmn,mps07,marquet07,  
buw,soyez,watt08}  suggests that these effects might   
become important in the energy regime probed by current colliders. This  motivated the generalization of  the 
saturation model  to two-photon interactions at high energies performed in Ref. \cite{Kwien_Motyka},   which has 
obtained a very good description of the data on the $\gamma \gamma$ total cross section, on the photon structure 
function at low $x$ and on the $\gamma^* \gamma^*$ cross section. The formalism used in 
Ref. \cite{Kwien_Motyka} is 
based on the dipole picture \cite{dipole}, with the  $\gamma^* \gamma^*$ total cross sections 
being described  by the interaction of two color dipoles, in which the virtual photons 
fluctuate into (For previous 
analysis using the dipole picture see, e.g.,  Refs. \cite{nik_photon,dona_dosch}). The main 
assumption made in 
\cite{Kwien_Motyka} is that the dipole-dipole cross section 
can be expressed in terms 
of dipole-proton cross section with the help of  the additive quark model.  
This is a strong assumption which deserves a more detailed analysis.  
This is our first goal in this paper. In particular, we propose a more sophisticated  
connection between   the  dipole - dipole and  dipole - proton scattering amplitudes. Our second goal is to compare 
the predictions obtained using \cite{Kwien_Motyka}  with those obtained  with  our approach. We also discuss the dependence 
of the results on the dipole  scattering amplitude. Here we make use of  the state-of-the art parametrization of the 
dipole scattering amplitude  \cite{bkrunning}. We compare the results with   the currently available experimental data and 
provide  estimates of the total cross sections and photon structure functions which will be measured in the future linear colliders.
It is important to emphasize that  in \cite{Kwien_Motyka} the cross sections were estimated  considering the GBW model 
\cite{GBW}, which is inspired on saturation physics, and during  the last years an intense activity in the area resulted  in  more 
sophisticated dipole - proton cross sections  \cite{bgbk,kw,kmw,fss04,fss06,kkt,iim,dhj,
gkmn, mps07, marquet07, buw,soyez,watt08}, which could be used 
to estimate the $\gamma \gamma$ cross sections. In  our study we also compare the predictions obtained using the solution of the BK equation with those from the phenomenological saturation model proposed in \cite{iim} with free parameters updated in   \cite{soyez}.

Before introducing the required formulas, some comments are in order. 
Firstly, in our study the heavy quark contribution is not 
included, since the solution of the BK equation used in our calculations  
\cite{bkrunning} has its free parameters fixed  disregarding 
the contribution of heavy quarks to the inclusive and longitudinal structure functions. 
Recently, this solution was improved by including the charm and bottom contributions to 
these observables, which have strong effects on the fit parameters \cite{aamqs}. 
As this solution  is 
not yet public, we postpone for a future study the discussion of heavy quark production 
in $\gamma \gamma$ interactions. For consistence we also consider the phenomenological 
saturation model proposed in \cite{iim} without the inclusion of heavy quarks. However, we 
also consider the updated version obtained in \cite{soyez}, where the free parameters 
were fixed  considering the more recent H1 and ZEUS data. 
Secondly, we are assuming in our study that fluctuations and correlations produced by the BK equation in the dilute regime (see, for instance, Refs. \cite{pom_loop}), where this equation reduces to the BFKL equation, can be neglected in the calculation of the 
dipole - dipole scattering cross section. This is a strong assumption. However, results obtained using a toy model in \cite{portugal}, 
indicate that these effects are tamed by saturation in the high-density regime  and by the running of coupling in the dilute 
regime. As in our model the scattering amplitude is given in terms of the solution of the BK equation including running coupling 
corrections, we expect  the contribution of fluctuations to be small.

\begin{figure}
\vspace{1.0cm}
%\begin{tabular}{cc}
\centerline{\psfig{figure=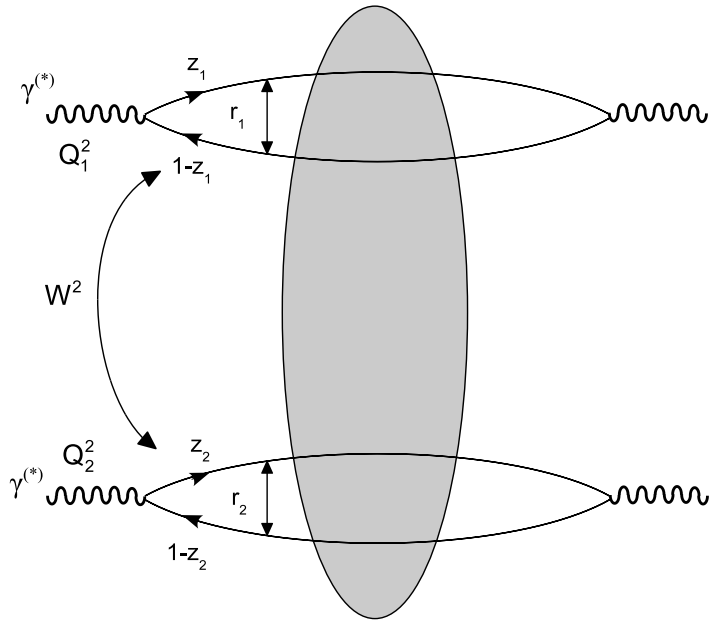,width=10cm}} 
%\end{tabular}
\vspace{0.5cm}
\caption{The diagram illustrating the $\gamma^* \gamma^*$ interaction in the
dipole representation. See formula (\ref{master}).}
\label{fig1}
\end{figure}

\section{The dipole picture for the two-photon cross section}

Let us start presenting a brief review of the two-photon interactions in the dipole picture.
At high energies, the scattering process can be seen
 as a succession in time of two
factorizable subprocesses (See Fig. \ref{fig1}): i) the photon fluctuates into  
quark-antiquark pairs (the dipoles), ii) these color dipoles interact and produce the final state.
The
corresponding cross section is given by
\be
\sigma_{ij}(W^2,Q_1^2,Q_2^2)\; = \; \sum_{a,b=1}^{N_f} 
\int dz_1 
\int    d^2 \rr_1 \, |\Psi_i^a(z_1,{\tdm r_1})|^2
\int dz_2
\int  d^2 \rr_2 \, |\Psi_j^b(z_2,{\tdm r_2})|^2
\; \sigma^{dd}_{a,b}(\rr_1,\rr_2,Y) ,
\label{master}
\end{equation}
where $W^2$ is the collision center of mass energy squared, 
the indices $i,j$ label the polarisation states of the virtual photons,
i.e. $T$ or $L$, ${\tdm r}$ denotes the transverse separation between $q$ and $\bar q$
in the color dipole, $z$ is the longitudinal momentum fraction of the
quark in the photon and $Y$ is the available rapidity interval,  $Y \approx \ln (W^2 / Q_1Q_2)$. The wave functions $|\Psi_T^f(z,\rr)|^2$ and $|\Psi_L^f(z,\rr)|^2$ are given by
\begin{eqnarray}
|\Psi_{T}^f(z,{\tdm r})|^2\; &  = & \;
{6\alpha_{em}\over 4 \pi^2} e_f^2\{
[z^2+(1-z)^2]\;\epsilon_f ^2 K_1^2 (\epsilon_{f}r)
+m_f^2\,K_0^2(\epsilon_{f}r)\}\,\,, 
\\
 |\Psi_{L}^f(z,{\tdm r})|^2\; & = & \;
{6\alpha_{em}\over 4 \pi^2} e_f^2
 \,[4 Q^2 z^2 (1-z)^2 \; K_0^2 (\epsilon_{f}r)]\,\,,\label{psit}
\end{eqnarray}
with  $(\epsilon_f)^2=z(1-z) Q^2 +
m_f^2$, 
$e_f$ and $m_f$ denote the charge and mass of the quark of flavor $f$ and the 
functions $K_0$ and $K_1$ are the McDonald--Bessel functions. 
Moreover,  $\sigma^{dd}_{a,b}(r_1,r_2,Y)$ are the
dipole-dipole total cross-sections corresponding to their
different flavor content specified by the $a$ and $b$ indices.

%\section{The dipole-dipole scattering cross section}

The main input for the calculation of the total cross section is the dipole - dipole cross section, which in the approximation of two gluon exchange between the dipoles is given by (See, e.g., appendix A from \cite{navelet})
\begin{eqnarray}
\sigma^{dd}(\rr_1,\rr_2,Y) = 2 \pi \alpha_s^2 r^2 [  1 + \ln (\frac{R}{r})]
\label{sigdd_el}
\end{eqnarray}
where $r  =$ Min$(r_1,r_2)$ and $R = $ Max$(r_1,r_2)$, and is energy independent.
In the  BFKL approach  the dipole - dipole cross section reads
\cite{dipole}
\begin{eqnarray}
\sigma^{dd}_{BFKL}(\rr_1,\rr_2,Y) = 2 \pi \alpha_s^2 r_1^2 \int \frac{d\gamma}{2 \pi i} \frac{(r_2/r_1)^{2\gamma}}{\gamma^2(1-\gamma)^2} \exp \left[ \frac{\alpha_s N_c}{\pi} \chi (\gamma) Y \right]
\label{sigdd_bfkl}
\end{eqnarray}
where $\chi$ is the BFKL characteristic function, which satisfies the property $\chi(\gamma) = \chi(1-\gamma)$. It implies that $\sigma^{dd}_{BFKL}$ is symmetric under the exchange $r_1 \longleftrightarrow r_2$ of the two dipoles. 

The behavior predicted by the BFKL equation implies that the cross section violates the unitarity at high energies. 
Consequently, unitarity corrections should be considered in order to tame the BFKL growth of the dipole scattering amplitude. In \cite{salam} this problem was addressed considering independent multiple scatterings between the onia within the color dipole picture, with unitarization obtained in a symmetric frame, like the center-of-mass frame. As demonstrated in \cite{iancu_mueller}, these unitarity corrections can also be estimated considering the Color Glass Condensate formalism, which provides a description of the non-linear 
effects in the hadron wavefunction. It is important to emphasize that in general the applications of the CGC formalism 
to scattering problems require an asymmetric frame, in which the projectile has a simple structure and the evolution occurs in the target wavefunction, as it is the case in deep inelastic scattering. Therefore the use of the solution of the BK equation in the 
calculation of the dipole - dipole scattering cross section is not a trivial task. Another aspect that deserves  a more 
attention  is related to the impact parameter dependence of the scattering amplitude. In the eikonal approximation the dipole - dipole cross section can be expressed as follows
\begin{eqnarray}
\sigma^{dd} (\rr_1,\rr_2,Y)  = 2 \int d^2\rb \,{\cal{N}}(\rr_1,\rr_2,\rb,Y)
\end{eqnarray}
where  ${\cal{N}}(\rr_1,\rr_2,\rb,Y)$  is the scattering amplitude for the two dipoles with transverse sizes $\rr_1$ and $\rr_2$, relative impact parameter $\rb$ and rapidity separation $Y$. The scattering amplitude ${\cal{N}}$ is related to the  $S$-matrix by $S = 1-{\cal{N}}$, with the unitarity of the $S$-matrix implying ${\cal{N}} \le 1$. This constraint is obeyed by the solution of the BK equation. However, the dipole - dipole cross section can still rise indefinitely with the energy, even after the black disk limit (${\cal{N}} = 1$) has been reached at central impact parameters. It occurs due to the non-locality of the evolution, that keeps expanding the gluon distribution in the target towards larger impact parameters. This radial expansion is expected to occur logarithmically with the energy, in agreement with the Froissart bound, which should be contrasted with the power like rising towards the blackness at fixed impact parameter. Following \cite{ikt}, we will assume that the radial expansion only affects the subleading energy dependence of $\sigma^{dd}$ and study the approach towards unitarity limit at a fixed value of the target size. Moreover, we will assume that only the range $b < R$, where $R =$ Max$(r_1,r_2)$, contributes for the dipole - dipole cross section, {i.e.} we will assume that ${\cal{N}}$ is negligibly small when the dipoles have no overlap with each other ($b>R$). Therefore, we propose that the dipole-dipole cross section can be expressed as follows
\begin{eqnarray}
\sigma^{dd} (\rr_1,\rr_2,Y)  = 2 \, {{N}}(\rr,Y) \int_0^R d^2\rb = 2 \pi R^2 {{N}}(\rr,Y)\,\,, 
\label{geral}
\end{eqnarray}
where ${{N}}(\rr,Y)$ is independent of the impact parameter and satisfies the unitarity bound.  
The explicit form of $\sigma^{dd}$ reads
\begin{eqnarray}
\sigma^{dd} (\rr_1,\rr_2,Y) =  2 \pi r_1^2 N(r_2,Y_2) \, \Theta(r_1 - r_2) + 2 \pi r_2^2 N(r_1,Y_1) \, \Theta(r_2 - r_1) \,\,,
\label{ourmodel}
\end{eqnarray}
where  $Y_i = \ln (1/x_i)$ and 
\begin{eqnarray}
 x_i = \frac{Q_i^2 + 4 m_f^2}{W^2 + Q_i^2}.
\label{xdef}
\end{eqnarray}

In contrast, in the phenomenological model proposed in \cite{Kwien_Motyka} the dipole-dipole cross section was assumed to be given by 
\begin{equation}
\sigma^{dd}_{a,b}(r_1,r_2,Y) = \sigma_0^{a,b}\,{{N}}(\rr_1,\rr_2,Y)
\label{sigmadd_mot}
\end{equation}
with $\sigma_0^{a,b} = (2/3) \sigma_0$, where $\sigma_0$ is a free parameter in the saturation model considered, fixed by fitting the DIS HERA data. This relation can be justified by the quark counting rule, as the ratio between  the number of constituent quarks in a photon  and the
corresponding number of constituent quarks in the proton.
 Moreover, it was assumed that 
${{N}}(\rr_1,\rr_2,Y) = {{N}}(\rr_{\rm\small  eff},Y=\ln(1/\bar x_{ab}))$, where 
\begin{equation}
r^2_{\rm\small  eff}\; = \;{r_1^2r_2^2\over r_1^2+r_2^2}\,\,\, \mbox{and}\,\,\, \bar x_{ab} \; = \;{Q_1^2 + Q_2^2 +4m_a^2+4m_b^2\over W^2+Q_1^2+Q_2^2}\,\,.
\label{reff}
\end{equation}

In what follows we will calculate the 
total $\gamma \gamma$, $\gamma^{*} \gamma^{*}$ 
cross-sections and  the real photon structure function $F_2^{\gamma}(x,Q^2)$
considering these two models for the dipole-dipole cross section. Before, in the next Section, we discuss in more detail  the scattering amplitude $N(r,Y)$ used in our calculations.

\section{The forward dipole scattering amplitude}

\begin{figure}
\vspace{0.5cm}
%\begin{tabular}{cc}
\centerline{\psfig{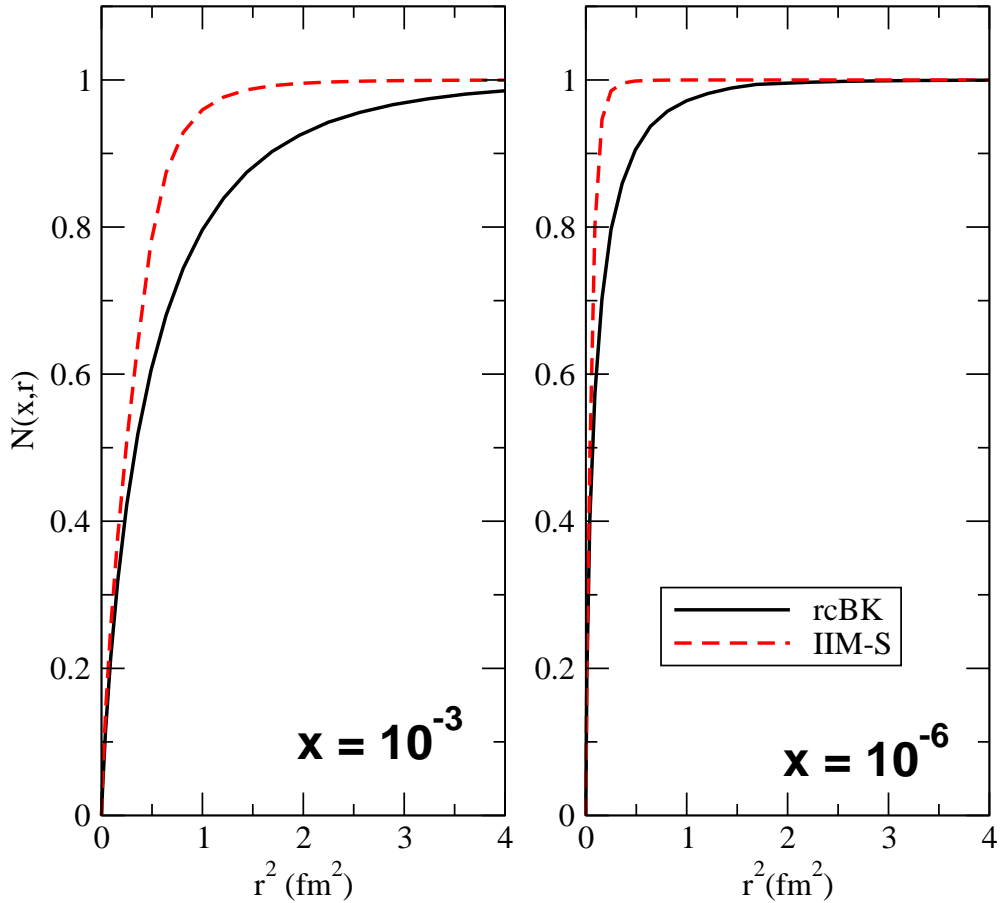}}
%\end{tabular}
\vspace{1.0cm}
\caption{Pair separation dependence of the rcBK (solid line) and IIM (dashed line) 
scattering amplitudes  at different values of $x$: (a) $x = 10^{-3}$ and (b) 
$x = 10^{-6}$.}
\label{fig2}
\end{figure}

The forward scattering amplitude ${\cal{N}}(x,\rr)$ is a solution of the  
Balitsky-Kovchegov (BK) equation, which is given in leading order by 
\begin{equation}
\label{eq:bklo} 
\frac{\partial {{N}}(r,Y)}{\partial Y} = \int {\rm d}\bm{r_1}\, 
K^{\rm{LO}}(\bm{r,r_1,r_2}) [{{N}}(r_1,Y)+{{N}}(r_2,Y)-{{N}}(r,Y)-
{{N}}(r_1,Y){{N}}(r_2,Y)],	
\end{equation}
where  $Y\equiv \ln(x_0/x)$ ($x_0$ is the value of $x$ where the evolution 
starts), and $\bm{r_2 = r-r_1}$. $K^{\rm{LO}}$ is the evolution kernel, given by 
\begin{equation}
\label{eq:klo}		
K^{\rm{LO}}(\bm{r,r_1,r_2}) = \frac{N_c\alpha_s}{2\pi^2}\frac{r^2}{r_1^2r_2^2},
\end{equation}
where $\alpha_s$ is the  strong coupling constant. This equation is a
generalization of the linear BFKL equation (which corresponds of the first three terms), 
with the inclusion of the (non-linear) quadratic term, which damps the indefinite growth 
of the amplitude with energy predicted by BFKL evolution. 
The leading order BK equation presents some difficulties when applied to study
DIS small-$x$ data. In particular, some studies concerning this equation
\cite{IANCUGEO,MT02,AB01,BRAUN03,AAMS05} have
shown that the resulting saturation scale grows much faster with increasing energy
($\lambda\simeq  0.5$ ) than that
extracted from phenomenology ($\lambda \simeq  0.2-0.3$). 
The calculation of the running coupling corrections to the  BK evolution kernel was explicitly
performed in \cite{kovwei1,balnlo}, where the authors included 
$\alpha_sN_f$ corrections to the kernel to all orders. In \cite{bkrunning} the 
improved BK equation was numerically solved replacing the leading 
order kernel  in Eq. (\ref{eq:bklo}) by the modified kernel which includes the running 
coupling corrections and  is given by \cite{balnlo} 	
\begin{equation}
\label{eq:krun}	 
K^{\rm{Bal}}(\bm{r,r_1,r_2})=\frac{N_c\alpha_s(r^2)}{2\pi^2}
		\left[\frac{r^2}{r_1^2r_2^2} + 
\frac{1}{r_1^2}\left(\frac{\alpha_s(r_1^2)}
		{\alpha_s(r_2^2)}-1\right)+\frac{1}{r_2^2}\left(\frac{\alpha_s(r_2^2)}
		{\alpha_s(r_1^2)}-1\right)\right].
\end{equation}
Numerical studies of the improved BK equation \cite{javier_kov} 
have confirmed that the running coupling corrections lead to a considerable 
slow-down of the evolution speed, which implies, for example, a slower growth of the 
saturation scale with
energy, in contrast with the faster growth predicted by the LO BK equation.
Since  the improved BK equation has been shown to be quite successful when applied to 
the description of the $ep$ HERA data for the proton structure function, we feel 
confident 
to use it  in other physical situations such as $\gamma  \gamma$ colisions. In what 
follows   
we make use of the public-use code available in \cite{code}.

The running coupling Balitsky - Kovchegov (rcBK) predictions will be compared with those
 from the parametrization proposed in Ref. \cite{iim} and updated in \cite{soyez}, which was  constructed so as 
to reproduce two limits  of the LO BK equation 
analytically under control: the solution of the BFKL equation
for small dipole sizes, $r\ll 1/Q_s(x)$, and the Levin-Tuchin law 
for larger ones, $r\gg 1/Q_s(x)$. In the updated version of this parametrization \cite{soyez}, the free parameters were obtained by fitting more recent H1 and ZEUS data.
In this parametrization the dipole forward scattering amplitude is given by
\begin{eqnarray}
{{N}}(x,\rr) =  \left\{ \begin{array}{ll} 
{\mathcal N}_0\, \left(\frac{r\, Q_s}{2}\right)^{2\left(\gamma_s + 
\frac{\ln (2/r Q_s)}{\kappa \,\lambda \,Y}\right)}\,, & \mbox{for $r 
Q_s(x) \le 2$}\,,\\
 1 - \exp^{-a\,\ln^2\,(b\,r\, Q_s)}\,,  & \mbox{for $r Q_s(x)  > 2$}\,, 
\end{array} \right.
\label{CGCfit}
\end{eqnarray}
where $a$ and $b$ are determined by continuity conditions at $\rr Q_s(x)=2$,   
$\gamma_s= 0.6194$, $\kappa= 9.9$, $\lambda=0.2545$, $Q_0^2 = 1.0$ GeV$^2$,
$x_0=0.2131$ and ${\mathcal N}_0=0.7$. Hereafter, 
we shall call the model above  IIM-S.
The first line from Eq. (\ref{CGCfit}) describes the linear regime whereas the 
second one describes saturation effects. 

In Fig. \ref{fig2}  we compare the pair separation dependence of the  rcBK and IIM-S forward dipole scattering amplitudes
  at distinct values of $x$. The main difference 
between these models is  the rapid 
onset of saturation predicted by the IIM-S model. In comparison,  the rcBK 
solution 
predicts a smooth growth, with a delayed saturation of the forward dipole scattering amplitude. Basically, the asymptotic saturation regime is only observed for very small values 
of $x$, beyond the kinematical range of HERA.

\section{Results and discussion}

\begin{table}[t]
\centering
\begin{tabular}{|c|c|c|c|}
\hline
Model & $N(\rr,Y)$   & $m \, (\mbox{MeV})$ & $\Lambda \, (\mbox{MeV})$   \\
\hline
Model 1 & rcBK &  198  &  ---   \\
\, & IIM-S  &  205  &  ---   \\
\hline
Model 2 & rcBK &  ---  &  210  \\
\,& IIM-S  &  --- &  230    \\
\hline
\end{tabular}
\caption{Parameters used in the calculations.}
\label{tab1}
%\end{center}
\end{table}

Before  presenting our results let us discuss the free parameters in our calculations. In the phenomenological model proposed in \cite{Kwien_Motyka}, denoted in what follows Model 1, the parameter which is adjusted in order to describe the experimental data of the total $\gamma \gamma$ 
cross section is the mass of the light  quarks. In order to describe the normalization of the data the authors are obliged to assume values which are larger  than those obtained in the fit of the $F_2$ HERA data using the same phenomenological 
saturation model.    Here we follow the same procedure and adjust the light quark mass in the wavefunctions in order to fit 
the real total cross section when using the rcBK and IIM-S models for the dipole scattering amplitude.
On the other hand, when using the dipole scattering amplitude from Eq. (\ref{ourmodel}), denoted Model 2 hereafter, we constrain the light quark mass to the values obtained in the original fits of the $F_2$ HERA data \cite{bkrunning,soyez}. However, in  Model 2, 
due to the quadratic dependence on the size of the larger dipole [See Eq. (\ref{geral})], the contribution of large values of $r_1$ and $r_2$ is quite significant in  the total cross section. In order to keep our calculations in the perturbative regime we cut the 
integration on the pair separation at  a maximum value of the order of the inverse perturbative QCD energy scale.  
In other words, we stop the $r_1$ and $r_2$ integrations at a maximum dipole  size, which 
is chosen to be  $r_{max}  =  \frac{1}{\Lambda}$, with $\Lambda$ a free parameter in the 
model which is expected to be $\approx\Lambda_{QCD}$. 
This parameter will be fitted in order to describe the total $\gamma \gamma$ cross section 
data at high energies. { In principle we could keep  $\Lambda = \Lambda_{QCD}$ fixed, understanding it as a clean-cut frontier 
between perturbative and non-perturbative physics. We could then compute the 
photon-photon cross sections, compare them with data and observe   how well perturbative QCD works in this domain. 
Discrepancies between theory and data would be attributed to non-perturbative contributions.  We expect these contributions to be 
larger in the case of real photons, which have a larger average transverse radius. However the uncertainties in the value of  
$\Lambda_{QCD}$ would make this separation between the perturbative and non-perturbative regimes  less precise. In the present work we 
adjust the value of  $\Lambda$. If the value  required to fit the data would be much different (e.g. much smaller) from  $\Lambda_{QCD}$ 
this would be an indication that it is not possible to describe the bulk of data only with perturbative QCD. Surprisingly the obtained 
values of $\Lambda$, shown in Table I, are close to the most accepted values of  $\Lambda_{QCD}$, suggesting that the physics of 
high energy photon-photon scattering is to a large extent perturbative.}

\begin{figure}[t]
%\begin{tabular}{cc}
\includegraphics[scale=0.3] {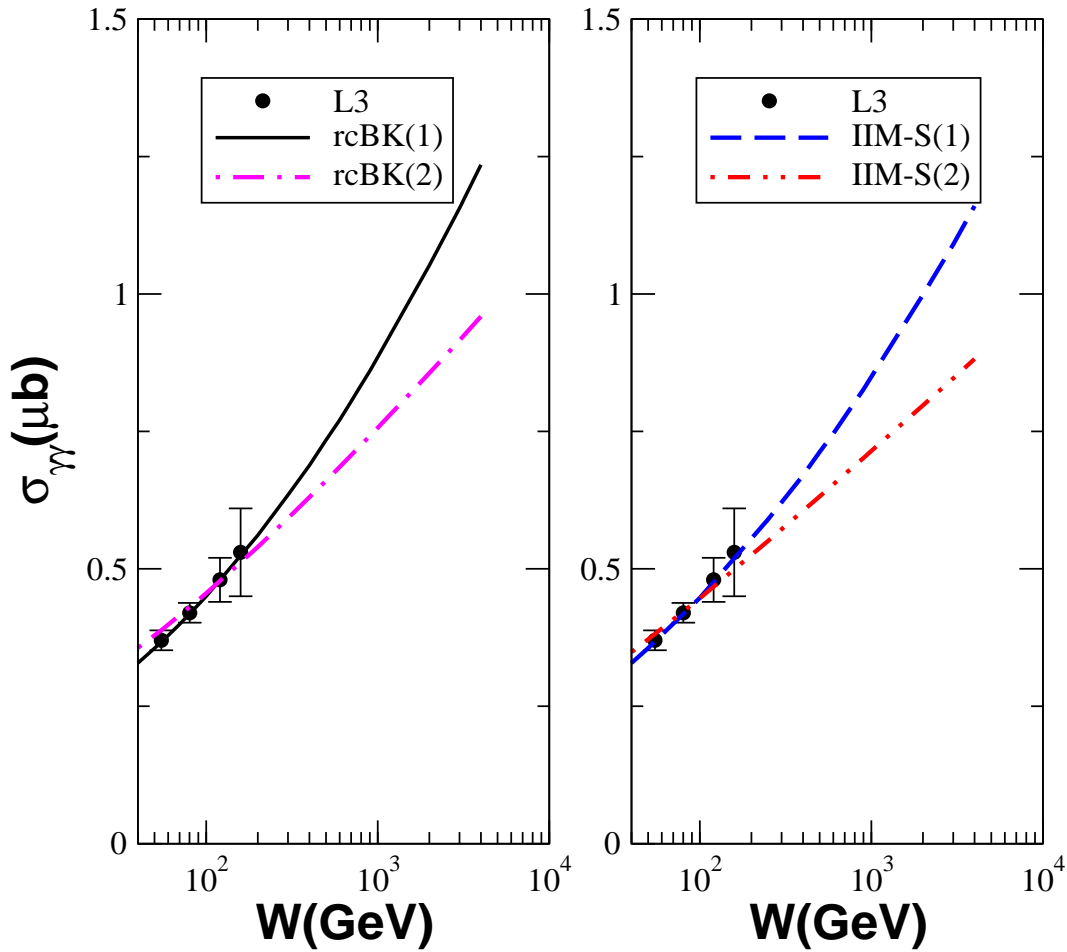}  
%%%%%\includegraphics[scale=0.45]{realdata3.eps} 
%(a) &  (b)
%\psfig{file=hqpp_charm_lhc.eps,width=83mm} & %\psfig{file=hqpp_bottom_lhc.eps,
%width=80mm}
%\end{tabular}
\vspace{0.5cm}
\caption{The total $\gamma \gamma$  cross section as a function of the energy $W$ for different models of dipole-dipole cross section and dipole scattering amplitude.  }
\label{fig3}
\end{figure}

In Fig. \ref{fig3} we present a comparison between the predictions obtained using the two models for $\sigma^{dd}$ and $N(r,Y)$ and the current experimental data for the  total $\gamma \gamma$ cross section at high energies, where for instance rcBK(1) indicates that we are using the rcBK solution for the scattering and the Model 1 for the dipole-dipole cross section and so on. It is important to emphasize that differently from \cite{Kwien_Motyka},  the Reggeon  contributions are not included in our calculations, since our focus 
is in the energy range $W \ge 50$ GeV, where these contributions are very small. 
The parameters used in our calculations are presented in Table \ref{tab1}. 
As in  \cite{Kwien_Motyka}, the description of the experimental data \cite{l3_real} using the Model 1 is 
only possible if we assume larger values of the light quark mass in comparison to those used in the description of the $F_2$ data, where $m_{u,d,s} = 140$ MeV.  On the other hand, the value of $\Lambda$ necessary to describe the experimental data is almost equal to $\Lambda_{QCD}$, in agreement with our expectations. This result can be interpreted as an indication that  Model 2 for the dipole - dipole cross section captures the 
main features  of the interaction. Finally, the predictions of the two models for $\sigma^{dd}$ are very similar in the kinematical range of the experimental data, independently of the $N(r,Y)$ considered. However, at $W > 110$ GeV, the predictions are very distinct. In particular, at $W = 1000$ GeV they differ by $\gtrsim 25$ \%, with  Model 2 predicting smaller values for the total $\gamma \gamma$ cross section.

\begin{figure}
\vspace{0.5cm}
%\hspace{-15.0cm}
%\begin{tabular}{cc}
\includegraphics[scale=0.3]{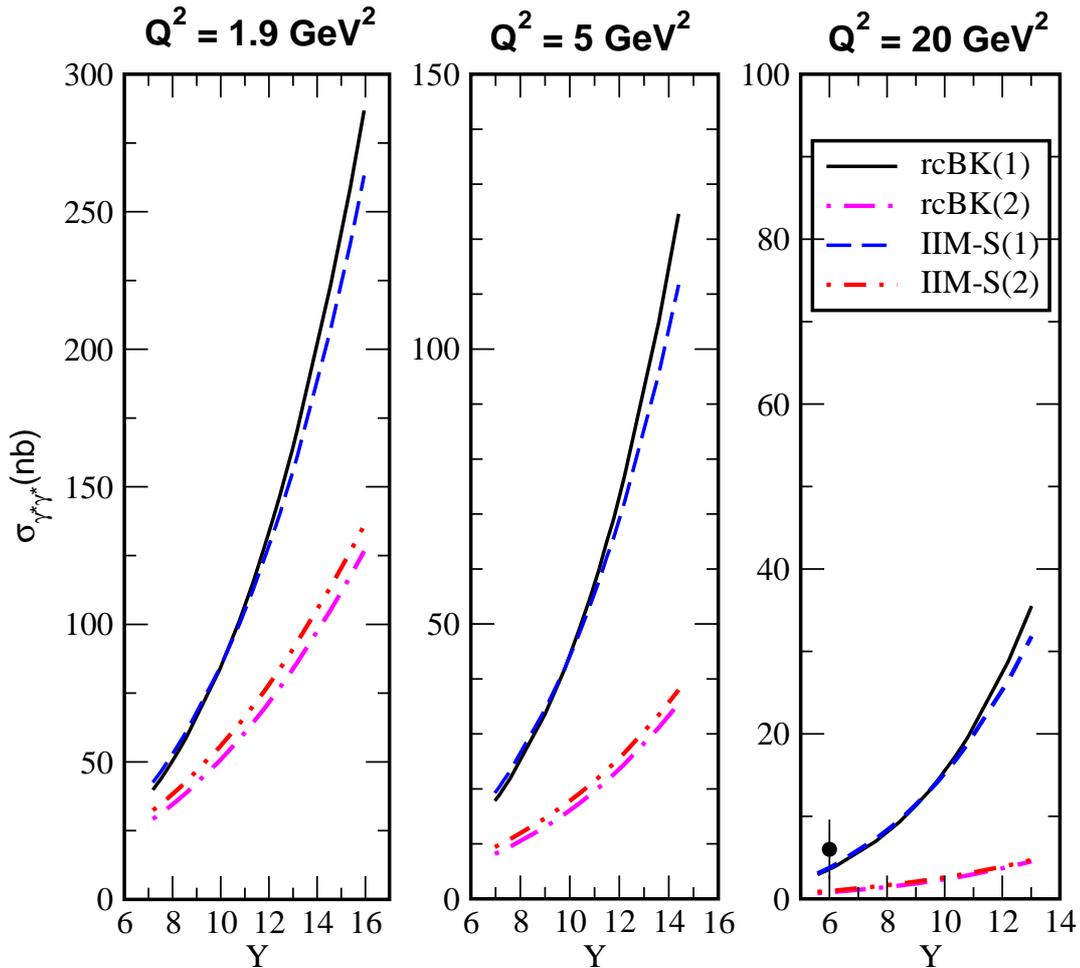}
%\centerline{\psfig{figure=gama_new_fig2.eps,width=12cm}} 
%\centerline{\psfig{virtualdata3.eps,width=12cm}} 
%\end{tabular}
\vspace{1.0cm}
\caption{The total $\gamma^*  \gamma^*$ cross sections as a function of the variable 
$Y\equiv \ln (W^2/Q_1 Q_2)$ for different values of $Q^2$ ($Q^2=Q_1^2=Q_2^2$).}
\label{fig4}
\end{figure}

We can also compute the two-photon cross section for the 
case $Q_1^2 \sim Q_2^2$ (with large $Q_{1,2}^2$) corresponding to the 
interaction  of two (highly) virtual photons and also for  the case $Q_1^2 \gg Q_2^2$  
corresponding to probing the structure of virtual ($Q_2^2 > 0$)  or real ($Q_2^2 =0$) photon at small values of the 
Bjorken  parameter $x=Q_1^2/(2q_1q_2)$  ($Q_i^2 \equiv - q_i^2$).  For instance, the structure function
 $F_2^{\gamma}(x,Q^2)$ of the real photon ($Q_2^2=0, Q_1^2=Q^2$)   is  related in the following way to the 
$\gamma^* \gamma$  total cross-sections:
\be
F_2 ^{\gamma}(x,Q^2)\; = \;
{Q^2\over 4 \pi^2 \alpha_{em}}
[\sigma_{T,T}(W^2,Q^2,Q_2^2=0) + \sigma_{L,T}(W^2,Q^2,Q_2^2=0)].
\label{fgg}
\ee
In Fig. \ref{fig4} we present our predictions for the total $ \gamma^* \gamma^*$ 
cross section for different photon virtualities. We assume that $Q_1^2 = Q_2^2 = Q^2$ 
and analyse the dependence of the cross section on the variable  
$Y\equiv \ln (W^2/Q_1 Q_2)$.  We can see that the cross sections increase with 
$Y$ and decrease with $Q^2$.  Moreover, similarly to the real case, the main difference between the predictions is 
associated to the choice of  $\sigma^{dd}$, with  Model 1 predicting larger values for the cross section and a steeper growth 
in  rapidity. This difference increases at larger values of the  photon virtuality, being a factor $\approx 8$ for $Y =10$ and $Q^2 = 20$ GeV$^2$. The experimental point in the right panel is from the L3 Collaboration \cite{l3_virtual}.

\begin{figure}
\vspace{0.5cm}
%\begin{tabular}{cc}
\includegraphics[scale=0.3]{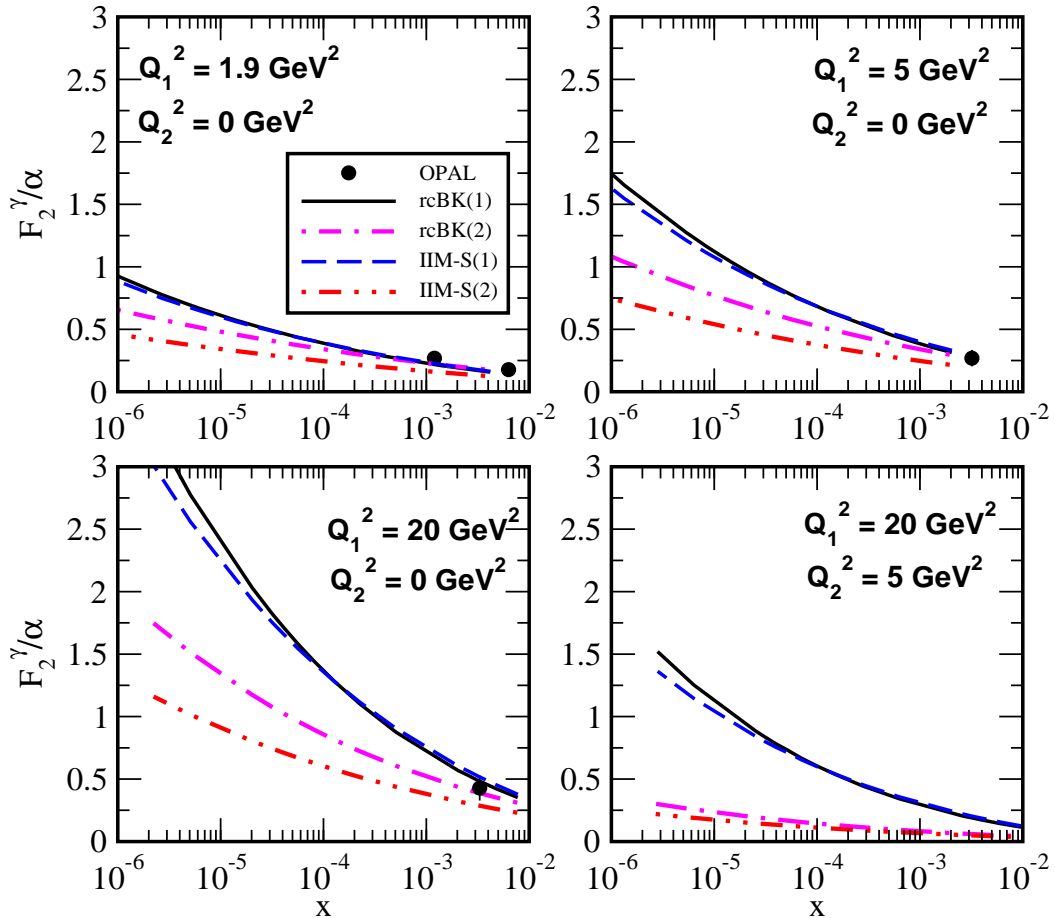}
%\centerline{\psfig{f2data3.eps,width=12cm}}
%\centerline{\psfig{figure=gama_new_fig3.eps,width=12cm}} 
%\end{tabular}
\vspace{1.0cm}
\caption{ The photon structure function $F_2^{\gamma}(x,Q^2)$ as a 
function of $x$ for different choices of the virtualities $Q_1^2$ and $Q_1^2$. }
\label{fig5}
\end{figure}

Finally, in Fig. \ref{fig5} we present our predictions for the $x$ dependence of the 
photon structure function $F_2^{\gamma}(x,Q^2)$ for different values of the  photon virtualities. The basic idea is that the quasi-real photon structure may be probed by other  photon with a large momentum transfer. We present in the lower right  panel our predictions for the virtual photon structure function. Although there exist only very few data on this observable, its experimental study is feasible in  future linear colliders. Our results predict that $F_2^{\gamma}(x,Q^2)$ increases at small-$x$, similarly to predictions for the  proton structure function.  The current experimental data \cite{f2_gama} are described quite well. 
As it was seen previously  in the case of $\sigma_{\gamma \gamma}$ and $\sigma_{\gamma^* \gamma^*}$,  Model 1 predicts a 
steeper growth with the energy and, consequently, at smaller values of $x$, with the difference between the models increasing at larger values of $Q^2$. An important aspect is that the predictions obtained using  Model 1 are almost independent of the scattering amplitude used in the calculations. On the other hand, in  Model 2, the predictions depend more 
strongly on $N(\rr,Y)$.  This makes the study of   $F_2^{\gamma}(x,Q^2)$ an 
important source of information about the QCD dynamics at high energies.

\section{summary}

In this paper we have estimated the main observables 
to be studied in $\gamma \gamma$ collisions in the future linear colliders 
using the color dipole picture. In this approach the main input is the dipole - dipole 
cross section, which is determined by the QCD dynamics. 
We have discussed  this quantity in detail and have introduced a new relation with the dipole scattering amplitude. In our calculations we use the 
state-of-the-art of the non-linear QCD dynamics, with the dipole  scattering amplitude 
given by the solution of the running coupling Balitsky-Kovchegov equation which are compared with the predictions of the phenomenological saturation model proposed in \cite{iim} and updated in \cite{soyez}. Moreover, we compare our predictions with those obtained using the phenomenological model for the dipole - dipole cross section proposed in \cite{Kwien_Motyka}. We demonstrate that these models are able to describe the scarce experimental data at currently available energies. However, they differ largely at higher  energies, which implies that future experimental data could be used to constrain the QCD dynamics.

\begin{acknowledgments}

This work was  partially financed by the Brazilian funding 
agencies CNPq and FAPESP.

\end{acknowledgments}

\hspace{1.0cm}

\end{document}